\documentclass[12pt]{article}

\usepackage{graphicx, wrapfig, subcaption, setspace, booktabs}
\usepackage{amsmath}
\usepackage{amssymb}
\usepackage{booktabs}
\usepackage{xcolor}
\usepackage{array}
\usepackage{multirow}
\usepackage{enumitem}
\usepackage{setspace}
\usepackage[margin=1in]{geometry}
\usepackage{lineno}
\usepackage{linenoaa}
\usepackage{ulem}
\usepackage{url}
\usepackage{natbib}
\bibpunct{(}{)}{;}{a}{}{,}
\newcommand\HI{H{\sc i}}
\newcommand{\secsp}{\vspace{-0.2cm}}

\begin{document}
\hspace{-1.3cm}
\includegraphics[width=1.1\textwidth]{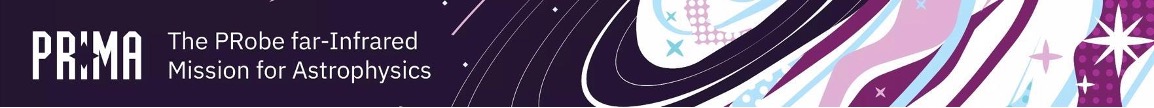}

\vspace{0.5cm}\hspace{-0.7cm}
\textbf{\large{Title}}: {Tracing Cloud Formation and Evolution Through 3D Magnetic Field Mapping}

\vspace{0.5cm} \hspace{-0.7cm}
\textbf{\large{First author}} (Affiliation): Mehrnoosh Tahani (University of South Carolina, Stanford University)

\vspace{0.5cm}\hspace{-0.7cm}
\textbf{\large{Co-authors}} (Alphabetized by last name):
Laura Fissel (Queen's University), Enrique Lopez Rodriguez (University of South Carolina), Kate Pattle (University College London)

\section*{Abstract}

We propose to use the unprecedented polarization sensitivity of PRIMA's PRIMAger Polarization Imager and its high resolution in Band 1 (92\,$\mu$m) to map magnetic fields across two contrasting molecular cloud environments: the well-studied Perseus cloud and the isolated Musca filament. This comparative study will leverage the existing VLA radio observations that provide line-of-sight magnetic field component  of the Perseus cloud, along with upcoming POSSUM survey results for Musca, to construct the first detailed 3D magnetic field vector maps at sub-parsec resolution. Perseus, with its known formation history through interstellar structure (e.g., bubble) interactions, will reveal how magnetic fields evolve during active star formation phases, while Musca, an isolated filament with lower star-formation activity, will show magnetic field morphology in early evolutionary stages. With PRIMA's resolution of 0.01\,pc for Perseus and $<0.01$\,pc for Musca, we will resolve magnetic field structures at scales necessary for understanding cloud fragmentation and star formation efficiency, and the roles that magnetic fields play in these processes. Our survey will require approximately 1438 hours to cover 42 deg$^2$ of Perseus and 12 deg$^2$ of Musca. With this, we aim to provide the first comprehensive view of how environment and evolutionary state may influence magnetic field evolution in molecular clouds and how magnetic fields influence cloud formation, fragmentation, and star formation.

% Perseus 2 MJy/Sr , 6*7 , 194.6 hours
% Musca 2 MJy/Sr , 4*3, 55.6 hours

% \textit{(0.5 page limit)}

\section*{\includegraphics[height=0.8em]{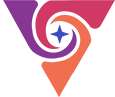}Science Justification}

\paragraph*{Unveiling Cloud History Through 3D Magnetic Fields:} Magnetic fields play significant roles in molecular cloud formation, evolution, and star formation \citep{Pattleetal2023PP7, PlanckXXXV}. However, understanding their true impact requires mapping three-dimensional (3D) field morphologies across multiple scales. Recent studies by \cite{Tahanietal2022O, Tahanietal2022P} mapped 3D interstellar magnetic field vectors for the first time, revealing the full orientation and direction of these vectors and providing the only complete 3D interstellar field reconstructions to date. Using these 3D vectors in 3D space (6D maps), they demonstrated how field morphology can reveal the formation history of molecular clouds, as an archaeological record of past interactions with galactic structures.

\textbf{Perseus -- Probing Magnetic Field Evolution During Active Star Formation:} The 3D field morphology of the Perseus cloud revealed a formation scenario inconsistent with the previously proposed Per-Tau supershell alone~\citep{Bialy2021}. The reconstructed magnetic field structure predicted the existence of an additional interacting structure, which was subsequently confirmed by kinematic observations \citep{Kounkeletal2022}. This success demonstrates the power of 3D magnetic field studies for understanding cloud formation and evolution.

These 3D reconstructions included both radio observations for Faraday rotation \citep{Tahanietal2018} and far-infrared observations by Planck \citep{Tahanietal2019, Tahanietal2022O, Tahanietal2022P, Tahani2022}. Due to resolution and sensitivity limitations of both datasets, the resulting 3D magnetic field vectors provide only coarse spatial information ($> 1$\,pc). To understand the detailed roles that magnetic fields play in cloud evolution, we need to improve the resolution of these 3D fields to sub-parsec scales. PRIMA's unprecedented resolution and sensitivity, combined with currently available and upcoming radio observations, make this advancement possible.

\begin{figure}[h!]
    \centering    
    \secsp
    \secsp
    \includegraphics[width=0.43\linewidth]{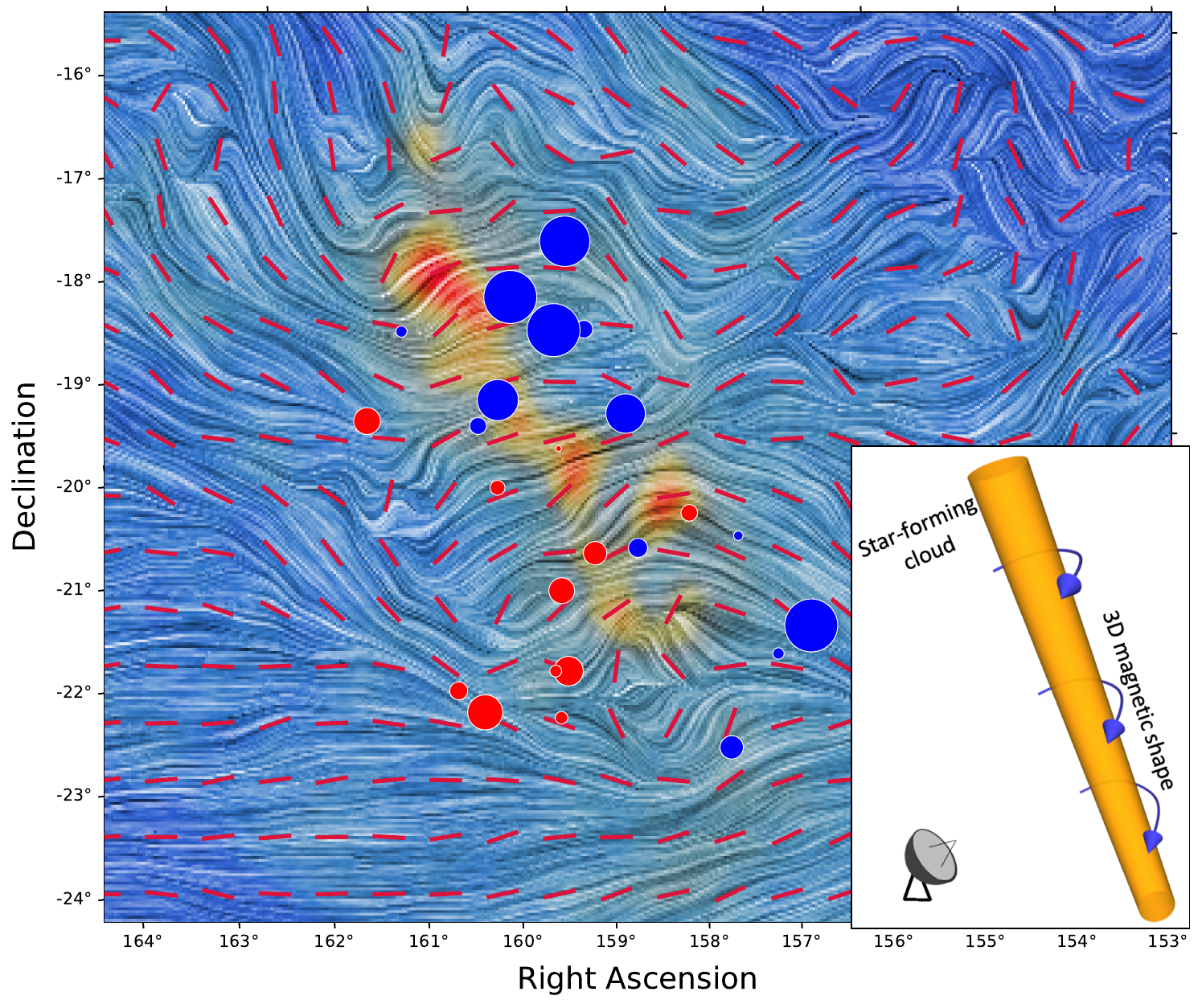}
    \includegraphics[width=0.5\linewidth]{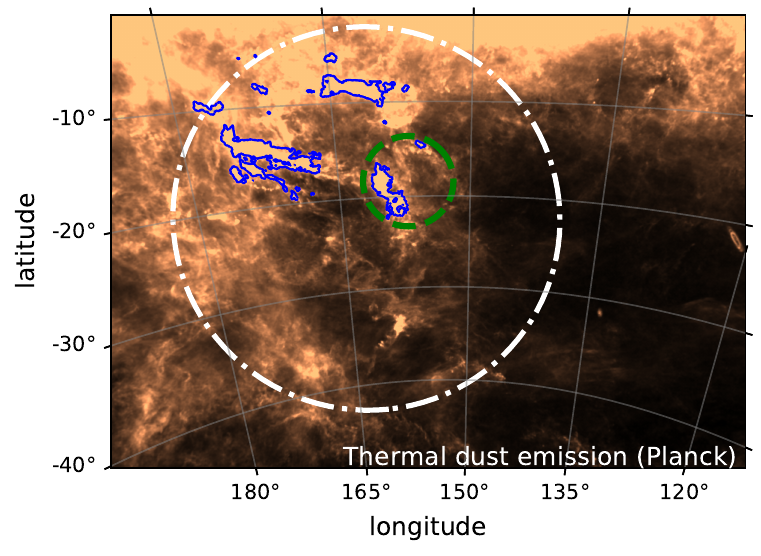}
    \vspace{-0.4cm}
    \caption{ \textbf{Left:} Perseus cloud's line-of-sight and 3D magnetic fields. Blue/red circles: line-of-sight magnetic fields toward/away from us. Red/drapery lines: Planck-observed plane-of-sky fields. Background: visual extinction map \citep{Kainulainenetal2009}. \textbf{Right:} Bubbles shaping Perseus (blue contour inside the green circle). Per-Tau (white) and newly-identified structure (green - shape not known) influence cloud. Background: thermal dust emission.}
    \label{fig:3D}
\end{figure}

Perseus represents an ideal region for studying magnetic field evolution during active star formation. The existing VLA L-band observations provide a high-density rotation measure measure map of the Perseus cloud (Hajizadeh et al. in prep), improving the cloud's Faraday rotation measure source density by a factor of several compared to previous studies. Additionally, upcoming rotation measure maps from the Polarisation Sky Survey of the Universe's Magnetism~\citep[POSSUM;][]{Jungetal2024, Gaensleretal2025}, SPICE-RACS~\citep{Thomsonetal2023SPICERACS}, and ultimately the full Square Kilometre Array will further improve line-of-sight magnetic field resolution and source density. Current plane-of-sky observations from Planck, however, lack the resolution to reveal magnetic field evolution at scales where gravity or turbulence may dominate cloud dynamics. PRIMA's capabilities will enable us to map 3D magnetic field morphology at sub-parsec scales both in the cloud environment and inside molecular clouds, trace magnetic field evolution from quiescent to active star-forming regions with varying stellar ages, and quantify magnetic field-density relationships in detail.

\paragraph*{Musca -- Magnetic Fields in Early Stages of Cloud Evolution:} 
%------------------------------------------
\begin{wrapfigure}{l}{6.65cm}
\vspace*{-.5cm}
\centering
\hspace*{-1.0cm}
\includegraphics[width=0.45\textwidth, trim={0cm 0.35cm 0 0.25cm},clip]{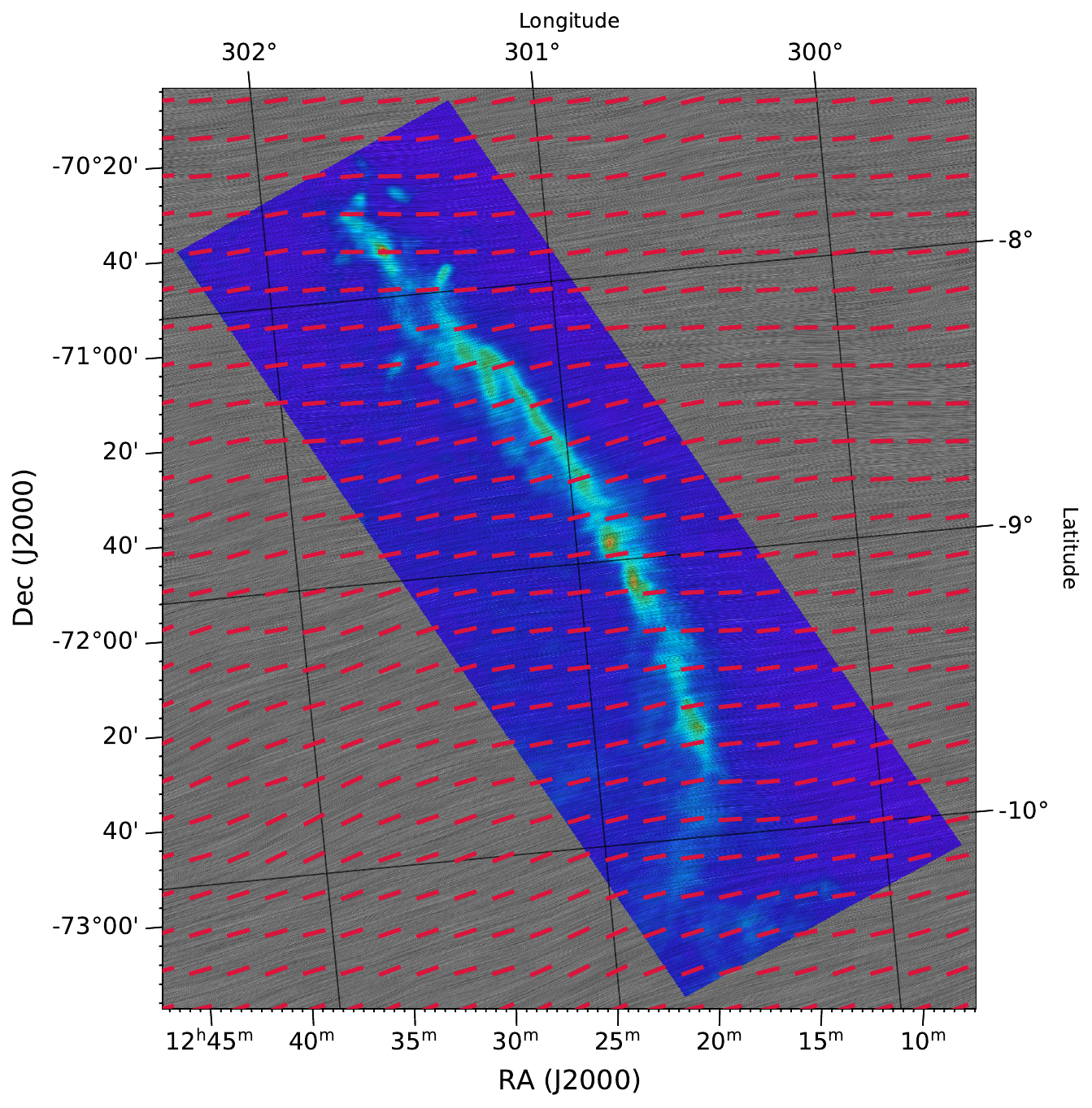}
\caption{Herschel (background color) and Planck observations of Musca. The red and drapery lines show the plane-of-sky magnetic fields.}
\label{fig:MuscaPOS}
\end{wrapfigure} 
\vspace*{-.5cm}
%------------------------------------------
The Musca filament offers significantly different properties compared to the Perseus cloud. Musca is an isolated, quiescent molecular cloud in the early stages of evolution \citep{Evansetal2014, Coxetal2016}. Located at $\sim$170\,pc distance \citep{Zuckeretal2021} with minimal active star formation, Musca represents magnetic field morphology largely unperturbed by stellar feedback (see Figure~\ref{fig:MuscaPOS}) with relatively high observed polarization fraction through dust polarization \citep{PlanckXIX2015, Ngocetal2021}. 
This contrasting environment is important for understanding several key aspects of magnetic field physics in molecular clouds, including: a) examining how magnetic fields influence filament formation in isolation, b) exploring the relationship between magnetic fields and cloud/filament fragmentation efficiency, c) investigating whether 3D magnetic field morphology differs fundamentally between bubble-compressed clouds like Perseus and isolated clouds like Musca, d) exploring the role of magnetic fields in setting initial conditions for star formation, and e) unveiling cloud-formation history through 3D magnetic field mapping. 
These insights will be necessary when combined with observations of a larger sample of clouds and their 3D field morphologies.

\paragraph{A Framework for Understanding Cloud Formation and Evolution}
The 3D magnetic field morphologies serve as unique tracers for deciphering the formation and evolutionary history of molecular clouds. This comparative study of Perseus and Musca will establish a framework for understanding how magnetic fields regulate molecular cloud evolution across different galactic environments and evolutionary stages. By combining PRIMA's sub-parsec resolution observations with  radio-derived line-of-sight magnetic fields~\citep{Tahanietal2025}, we will quantify for the first time how environmental factors such as bubble interactions versus isolation, and evolutionary states from active to quiescent star formation, influence magnetic field morphology, strength, and coupling with gas dynamics. This approach will enable us to explore cloud formation scenarios in different galactic environments and examine the role these scenarios play in the co-evolution of clouds and their magnetic fields. PRIMA's high sensitivity and spatial resolution will allow us to trace magnetic field evolution from large-scale galactic fields through cloud formation to gravitational collapse onset. This examination will provide observational constraints for magnetohydrodynamic simulations and theoretical models of star formation. This dual-cloud approach will serve as a foundation for future systematic studies of the broader molecular cloud population.

\section*{\includegraphics[height=0.8em]{prima-ident-small_edited.png}Instruments and modes used}

\begin{center}
\begin{tabular}{|l|l|l|}
\hline
\multicolumn{3}{|c|}{\textbf{PRIMAger}} \\
\hline
\centering
\textbf{Mapping details} & \textbf{Hyperspectral band} & \textbf{Polarimeter band} \\
& (24-84\,$\mu$m; R=8-10) & (96, 126, 172, 235\,$\mu$m; R=4) \\
\hline
Perseus: 6.0$^{\circ}$ × 7.0$^{\circ}$ & Not required & Yes \\
(1 map at 92\,$\mu$m) & & \\
\hline
Musca: 4.0$^{\circ}$ × 3.0$^{\circ}$ & Not required & Yes  \\
(1 map at 92\,$\mu$m) & & \\
\hline \hline
\end{tabular}   
\end{center}

% \textbf{If you selected Polarimeter Band, do you need polarimetry information?} Yes, for all bands

\paragraph{Approximate integration time:}
Our sensitivity requirements are driven by the need to detect polarized emission from diffuse cloud material and map magnetic field morphology at scales comparable to the radio observations.

\textbf{Perseus:} We target 5-$\sigma$ polarization detections for dust intensities of 20\,MJy/sr, based on Herschel Gould Belt observations, assuming 3\% polarization efficiency. This requires detection of 600\,kJy/sr polarized emission across our 6$^\circ$ × 7$^\circ$ (42\,deg$^2$) survey area. Using the PRIMA ETC for PPI Band 1 (92\,$\mu$m), this requires 972.6 hours of exposure time.

\textbf{Musca:} For the same sensitivity target of 20\,MJy/sr dust intensity with 3\% polarization fraction, we require detection of 600\,kJy/sr polarized emission. The 4$^\circ$ × 3$^\circ$ (12\,deg$^2$) survey area covering the main filament and surrounding regions requires 277.8 hours exposure time.

\textbf{Total time:} 1250.4 hours exposure time. Including 15\% overhead, the total requested time is \textbf{1438 hours}.

\paragraph{Special capabilities needed:}
N/A

\section*{\includegraphics[height=0.8em]{prima-ident-small_edited.png}Synergies with other facilities}

\paragraph*{Radio Observations:} Our study leverages existing and upcoming radio facilities to provide the line-of-sight magnetic field component required for 3D reconstruction. For Perseus, we will utilize completed VLA L-band observations providing high-density rotation measure data across the cloud (VLA/24A-376, VLA/20A-165, VLA/19B-053). These observations significantly improve source density compared to previous studies. For Musca, we will utilize upcoming rotation measure maps from POSSUM, SPICE-RACS, and ultimately the full Square Kilometre Array to obtain the first line-of-sight magnetic field measurements of this isolated filament. These studies incorporate the MC-BLOS technique \citep{Tahanietal2018, Tahanietal2025}, which uses Faraday rotation of background point sources combined with extinction maps and chemical evolution modeling to determine line-of-sight magnetic fields of molecular clouds. Additionally, targeted Zeeman splitting observations of OH and \HI\ lines will provide independent magnetic field strength measurements to complement our Faraday rotation results.

\paragraph*{Optical Polarimetry:} Complementary plane-of-sky stellar polarization observations from optical polarimetry surveys such as Dragonfly Polarimetry\footnote{\url{https://dragonflypol.github.io/DragonflyPol/}} will enable detailed mapping of foreground and background magnetic field contributions, essential for constructing accurate 3D magnetic field vectors of both clouds. These optical observations will help disentangle the complex line-of-sight structure by providing independent constraints on magnetic field orientations at different distances along the sight lines. Combined with Gaia stellar parallax measurements, this multi-wavelength approach will allow us to construct a complete 3D picture of magnetic field morphology from larger cloud scales down to individual star-forming cores and provide unprecedented insight into the magnetic field environment both within and surrounding the molecular clouds.

\paragraph*{Sub-mm Polarimetry:} Musca also will be mapped by the CCAT/FYST 6-m telescope in linear polarization at 350\,$\mu$m, 850\,$\mu$m, and 1.1\,mm using the PrimeCam instrument \citep{CCATPrime2021}.  With this PRIMAger + CCAT multiscale data we will be able to search for regions where there are differences in the inferred magnetic field direction as a function of wavelength. These differences could indicate that the magnetic field morphology changes for dust grain populations of different temperatures along the same sight-line through the cloud. This will allow us to probe magnetic field orientation as a function of dust temperature, complementing our 3D field measurements.  As ancillary science, with CCAT+ PRIMAger we will measure the dust polarization spectrum of Musca from 96\,$\mu$m to 1.1$\,$mm. We will compare our measurements with existing polarization spectrum predictions from different dust models to constrain the composition of the dust grain populations in Musca \citep{Guilletetal2018,Hensleyetal2023}.

\section*{\includegraphics[height=0.8em]{prima-ident-small_edited.png}Survey Design}
Our observational strategy employs PRIMA's PPI Band 1 (92\,$\mu$m) to map magnetic field morphology of the Perseus (6$^\circ$ × 7$^\circ$) and Musca (4$^\circ$ × 3$^\circ$) clouds. These observations will provide sub-parsec resolution plane-of-sky magnetic fields of both the cloud complexes and their surrounding environments. We will combine PRIMA's plane-of-sky magnetic field measurements with the radio-derived line-of-sight fields using established 3D reconstruction techniques. This approach will deliver the first comprehensive 3D magnetic field maps at sub-parsec scales for both clouds. These 3D maps serve as novel tracers of cloud formation and evolution~\citep{Tahanietal2022P, Tahanietal2022O} and will enable us to assess how environment and evolutionary state influence magnetic field and cloud co-evolution and star formation regulation. This study will help establish a framework for future systematic studies of the broader molecular cloud population.\footnote{claude.ai was used for editorial purposes.}

\bibliographystyle{aa} 
\bibliography{AllBiblio}

\end{document}